\documentclass{aa}
\usepackage{graphicx}

\def\nodata{ ~$\cdots$~ }

\begin{document}

\thesaurus{11 
	   (11.19.4;  
	    11.06.1;  
            11.09.1: NGC~5128)}  

\title{Deep VLT search for globular clusters in NGC~5128: color-magnitude
diagrams and globular cluster luminosity function
  \thanks{Based on observations collected at the European Southern 
    Observatory, Paranal, Chile, within the Observing Programme 63.N-0229,
	and at La Silla Observatory, Chile.}}

\author{Marina Rejkuba\inst{1,2}
}

\offprints{M. Rejkuba}

\institute{European Southern Observatory, Karl-Schwarzschild-Strasse
           2, D-85748 Garching, Germany\\
           E-mail: mrejkuba@eso.org
  \and Department of Astronomy, P. Universidad Cat\'olica, Casilla
        306, Santiago 22, Chile\\
}

\date{Received date / Accepted date}

\titlerunning{VLT search for globular clusters in NGC~5128}
\maketitle

%
%

\begin{abstract}
At the distance of NGC~5128 (3.6$\pm$0.2 Mpc) it is
possible to resolve globular clusters 
with high resolution imaging from the ground, 
thus allowing the globular cluster candidate selection primarily
through their morphological properties. 
I report the discovery of 71 globular clusters in NGC~5128 on VLT UT1+FORS1
images, including the faintest members (M$_V\sim-5$)
known to date in this galaxy as well as 5 previously known clusters. 
U- and V-band photometry has been measured for all the candidates and the
luminosity function, spanning $-10.1<M_V<-4.9$
 and $-9.3<M_U<-3.3$, constructed. 
These are the deepest globular cluster luminosity functions in an
elliptical galaxy determined so far. 
The Kolmogorov-Smirnov statistics show that the
difference between the globular cluster luminosity functions 
of NGC~5128 and the MW is not larger than 
the difference between the ones of M31 and the MW.
The (U$-$V)$_0$ color histogram shows a bimodal distribution. 
For 23 globular clusters I obtained K-band 
images with SOFI at the NTT in La Silla. Their positions 
in the (U$-$V) vs.\ (V$-$K) color-color diagram 
indicate that they are indeed old globular clusters.
Assuming that the globular clusters in NGC~5128 span a
similar age range as the ones in the Milky Way and adopting a linear fit
between the metallicity and (U$-$V)$_0$ color, the metal--rich clusters 
peak at [Fe/H]$=-0.6$ dex and the metal--poor ones peak at [Fe/H]$=-1.7$ dex. 
\keywords{Galaxies: star clusters -- Galaxies: formation -- 
          Galaxies: individual: NGC~5128}
\end{abstract}

\maketitle
%
%

\section{Introduction}

Globular Clusters are among the oldest objects in the Universe.
Moreover, they are intrinsically luminous and found around almost all
galaxies in which searches have been made. 
There are growing evidences for globular cluster
formation during
major star formation episodes (Schweizer \cite{schweizer}; Larsen \&
Richtler \cite{lr99}).
This makes them useful tracers of the formation and the evolution of
galaxies (Harris \cite{harris91,harris2000}; Kissler-Patig {\it et al.}\
\cite{kissler3}; van den Bergh \cite{vandenbergh}). 

The globular cluster luminosity function (GCLF) of a galaxy is the
relative number of its globular clusters per luminosity (or magnitude)
interval. It is usually unimodal and nearly symmetric with a peak at a
characteristic (turn--over) magnitude. 
Empirically it was found that turn--over magnitudes for
galaxies at the same distance (e.g.\ in the same galaxy cluster) 
differ by
$<0.1$, which makes the GCLF 
a good standard candle (Harris~\cite{harris91}; 
Jacoby {\it et al.}\ ~\cite{jacoby92}). However, 
the full calibration of the GCLF 
is incomplete. The essential problem is that the galaxies at large
distances, for which the GCLF method is mostly used, are
giant ellipticals in which globular clusters are found in largest numbers
and in which there are usually fewer problems with extinction, 
whilst the primary calibration of the standard candle $M_V^0$ rests on the
MW and M31, which are spiral galaxies. In fact, evidence that
GCLFs depend on Hubble type 
and environment has been claimed (Fleming {\it et al.}\ \cite{fleming}), but 
the former is probably due to differences in mean metallicity 
of the globular clusters and the latter
was never reliably demonstrated (Kissler-Patig \cite{kissler00}).

There is strong observational evidence that most luminous giant
ellipticals have a bimodal globular cluster
 metallicity distribution (Gebhardt \& Kissler-Patig 
\cite{gebh99}; Kundu \cite{kundu}). 
The globular cluster specific frequencies (the
number of clusters normalized to the galaxy luminosity; Harris \& van den
Bergh \cite{hvdenb81}) are
typically higher in ellipticals than in spirals (Harris 
\cite{harris91,harris2000}). 
To explain the high number of
globular clusters 
and the presence of several populations indicated by the bimodal
metallicity distributions, different globular cluster formation mechanisms
have been proposed (e.g.\ Ashman \& Zepf~\cite{ashmanzepf92}; C\^ot\'e {\it 
et al.}\ \cite{cmw98}; Forbes {\it et al.}\ \cite{forbes97}). 
The number, chemical abundance and spatial distribution of a globular
cluster system (GCS) put strong constraints on the formation of
 its parent galaxy.

At the distance of 3.6 Mpc (Soria {\it et al.}\ \cite{soria}; Harris {\it
et al.}\ \cite{hetal99}) NGC~5128 is the closest giant elliptical galaxy.
The first globular cluster in NGC~5128 was discovered by Graham \&
Phillips~(\cite{GP80}). Its brightness was high with $V=17.2\pm0.1$
mag, which corresponds to $M_V=-10.6$ mag.
Subsequent investigation of photographic plates (van den Bergh {\it et al.}\
~\cite{vBHH}, Hesser {\it et al.}\ ~\cite{HHvBH},~\cite{hetal86}) 
revealed a rich GCS in this
galaxy. The estimated number of globular clusters in NGC~5128 is
N$\sim$1700 
(Harris {\it et al.}\ \cite{hetal84})
compared with N$\sim$150 for the MW.

On the basis of
the near IR colors for a magnitude--limited sample of 12 globular clusters, 
Frogel~(\cite{Frogel80,Frogel84}) found one third of the sample to be 
super metal
rich (above solar metallicity) and luminous. However, the Washington
photometry of 62 spectroscopically confirmed clusters (Harris {\it et al.\
}~\cite{hetal92}) as well as a spectrophotometry for 5 clusters 
(Jablonka {\it et al.\ }~\cite{jablonka}) showed 
less extreme metallicity. 
Harris {\it et al.\ }~(\cite{hetal92}) found a large spread in
metallicities with a mean of [Fe/H]$_{C-T_1}=-0.8\pm0.2$ dex. 
They also find evidence
for a weak spatial metallicity gradient, but state that the same
could be produced by unknown reddening effects.
Zepf \& Ashman (\cite{zepfashman93}), using the same photometric data, 
suggest that the metallicity
distribution of globular clusters
 in NGC 5128 is bimodal, with the higher metallicity peak
at [Fe/H]=+0.25 dex, and proposed that this was 
due to the formation of a second population of globular clusters in
a previous merger which created NGC~5128.

Several photometric studies in the inner regions of NGC 5128 have been
conducted, with the aim of constraining the dependence of globular cluster
 metallicity on
galactocentric radius and of finding super-metal-rich clusters
(Minniti {\it et al.}\ \cite{minniti96}; Holland {\it et al.}\ 
\cite{holland}). 
These studies suggest the existence of radial
gradients and bimodal distribution in globular cluster
 metallicity. However, their
conclusions are limited
due to the restricted samples, possible
reddening within NGC~5128 that is especially severe in the
central parts around the prominent dust lane,
and possible contamination by stars
within NGC~5128 as well as background galaxies.

In this paper, I present a sample of globular clusters from two
fields in the halo of NGC~5128. Deep images combined with high resolving
power of the VLT and the proximity of NGC~5128 allow the identification of 
globular clusters primarily on the basis of their non-stellar PSF. 
Constructing the GCLF for a statistically significant number of clusters 
that span the whole range of magnitudes, I pursue the question of the
uniqueness of GCLF in elliptical galaxies. 

This paper is organized as follows. The data, 
the photometric calibrations and
completeness tests are described in Sect.~\ref{data}. In Sect.~\ref{CMD} 
the color--magnitude
diagram for the NGC~5128 globular clusters 
is compared to the ones of the MW and M31.
The luminosity function is presented in Sect.~\ref{LF} and the color
distribution in Sect.~\ref{color}. In the latter, the 
mean abundances are determined for the metal-poor and
metal-rich populations. The results are summarized
in Sect.~\ref{conclusion}.

%
%

 \section {The Data}
\label{data}

\begin{table*}
  \caption[]{Journal of Observations}
    \label{obslog}
      \begin{tabular}{lllllllll}
	\hline \hline
Field & $\alpha_{(2000)}$ & $\delta_{(2000)}$ & Date  
& Telescope \&    & Exposure & FILTER & Airmass&Seeing    \\
\#    & (h min sec)       & ($^\circ$  $\arcmin$  $\arcsec$)&  dd/mm/yy 
& Instrument &  (sec)   &        &        & $\arcsec$\\
\hline \hline

 1&13 26 23.5&-42 52 00&12/Jul/1999&Antu+FORS1  &2$\times$900&U & 1.536 & 0.52 \\ 
 1&13 26 23.5&-42 52 00&12/Jul/1999&Antu+FORS1  &2$\times$900&V & 1.768 & 0.54 \\ 
 2&13 25 24.0&-43 10 00&11/Jul/1999&Antu+FORS1  &2$\times$900&U & 1.283 & 0.53 \\ 
 2&13 25 24.0&-43 10 00&11/Jul/1999&Antu+FORS1  &2$\times$900&V & 1.196 & 0.44 \\
 2&13 25 24.0&-43 09 04&20/Feb/2000&NTT+SOFI    &45$\times$6$\times$10&K$_s$&1.055&0.56 \\
 1&13 26 48.1&-42 46 00&28/Feb/2000&2.2m+WFI    &3$\times$210&V & 1.117 & 1.00 \\
 2&13 26 48.1&-43 16 01&28/Feb/2000&2.2m+WFI    &3$\times$210&V & 1.037 & 0.81 \\
 2&13 24 07.6&-43 15 58&29/Feb/2000&2.2m+WFI    &3$\times$210&V & 1.057 & 1.12 \\
\hline
	\end{tabular}
\end{table*}

\subsection{Observations and Data Reduction}

I used the VLT Antu (UT1)+FORS1 (FOcal Reducer/low dispersion Spectrograph) 
to obtain the Bessel U- and V-band photometry of globular
clusters and stars in two different fields in NGC~5128. 
The field of view was $6\farcm8 \times 6\farcm8$ and
the pixel scale $0\farcs2/$pixel. The FORS1 detector is a
2048$\times$2048 CCD with 24$\mu$m pixels. 

Field~1 
was centered on the prominent N-E shell $\sim$14$^\prime$ away from the
center of the galaxy, while Field~2 
was chosen to overlap with 
Soria {\it et al.}\ (\cite{soria}) HST observations at a 
distance of $\sim$9$^\prime$ from the
center of the galaxy. Observations were carried out on the 11th and 12th
of July 1999 in service mode. Pairs of images of 15 min exposure time were
taken per field in each filter. The journal of observations 
is given in Tab.~\ref{obslog}.

In addition to VLT data, I obtained a 45 min long exposure of Field~2 
in K$_s$ using SOFI at the 
New Technology Telescope (NTT) at the ESO/La Silla 
Observatory. The field of view of SOFI is $4\farcm94 \times 4\farcm94$ 
and the pixel scale $0\farcs292/$pixel. This field is smaller than the one 
of FORS1 and so only about 40\% of objects in one field 
have K$_s$-band magnitudes. 

Both fields were also observed with the Wide Field Imager (WFI) at the
2.2m telescope on La Silla. 
The data from WFI observations are used here to 
assess the V-band magnitude of some of 
the brightest cluster candidates that were 
saturated on V-band FORS1 images. I calibrated the WFI photometry by 
comparing magnitudes of globular clusters well exposed (but not saturated) 
on VLT and WFI images in order
to determine the zero point of WFI observations.

For optical data (VLT and 2.2m images),
the standard image processing, including the overscan and 
bias subtraction and the
flat-field correction, was performed within the IRAF\footnote{IRAF is
distributed by the National Optical Astronomy Observatories, which is
operated by the Association of Universities for Research in Astronomy, Inc.,
under contract with the National Science Foundation} environment.
The pairs of exposures in each filter were registered using the {\em
imalign} task and averaged to obtain the
final images. Cosmic rays were rejected in this process using the {\it 
crreject}
algorithm within the {\em imcombine} task in IRAF. Bad pixels were masked out.

The standard procedure in reducing infrared (IR) data consists of 
sky subtraction, flat-field correction, registering and combining the images. 
Good sky subtraction in a crowded field like that of a galactic halo is 
particularly important. For that step I used the 
DIMSUM package (Stanford {\it et al.}\ \cite{dimsum}) within IRAF. 
In DIMSUM the sky subtraction is made in two passes. In the first one a median
sky is computed for each image from 6 temporarily closest frames.
The shifts between the sky subtracted frames are then computed and all the 
images stacked together using a rejection algorithm to remove cosmic rays. 
An object mask is computed for the coadded image and then shifted back in 
order to create object masks for the individual frames. In the second pass,
the sky subtraction is made using the object masks to avoid overestimation of 
the sky level. These masks are also used to check that the bright object
cores were not removed as cosmic rays in the previous pass. 
After the mask-pass sky subtraction, the frames are 
registered with {\em imalign} and combined with {\em imcombine} task in IRAF.

\subsection{Cluster identification and photometry}

\begin{figure}
   \resizebox{\hsize}{!}{\includegraphics{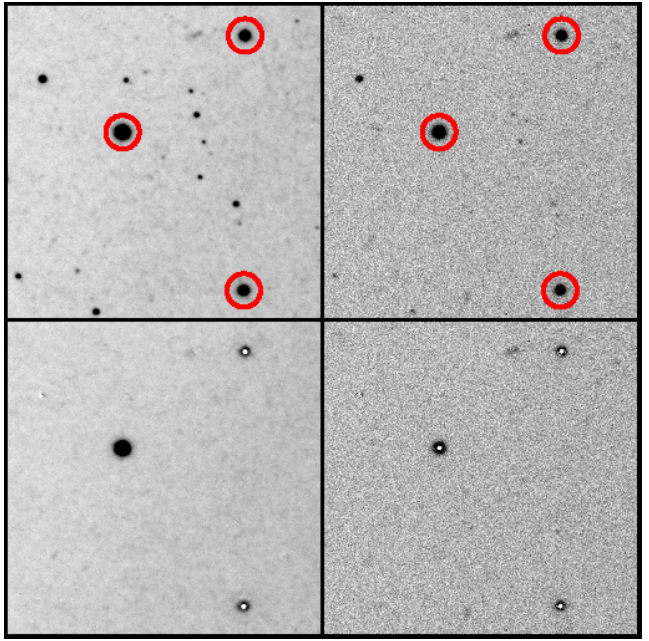}}
   \caption[]{Residual image of a $51\arcsec \times 51\arcsec$
portion of Field~2. On the left panels the
V-band image (top) and its residual image (bottom) and on the
right panels the corresponding U-band images are shown. Three globular
cluster candidates
are circled. The left-most globular cluster is slightly
overexposed in the V-band image. Its classification as a globular cluster
follows from the U-band residuals.}
   \label{residuals}  
\end{figure}

NGC~5128 is close enough that its globular clusters can be distinguished
from stars on the basis of their slightly non-stellar appearance on
high resolution ground-based images taken in excellent seeing conditions. 
At the distance of NGC~5128 (3.6 Mpc) a typical MW globular cluster
with a mean King core
radius of $r_c=2.3$ pc had, on the best seeing VLT images, a 
FWHM of 2.6 pix, which is $\sim0.4$ 
pix bigger than the stellar FWHM.
The excellent seeing, especially on the V-band images of Field~2, 
allowed me to identify globular cluster candidates
with $r_c$ as small as 1.7 pc, corresponding to observed FWHM of 2.4
pix (see f2.GC-6 in Tab.~\ref{nogcdata}), 
purely on the basis of their non-stellar
appearance through the following procedure within IRAF: (i) first the
{\em daofind} algorithm was run in order
to detect all the stars and clusters in the image; (ii) photometry
through a 3 pixels aperture was performed on all the objects; (iii)
relatively bright, isolated, stellar objects were used to create the PSF,
which was then fitted to all the objects, subtracted and the residual images
created. DAOPHOT~II (Stetson~\cite{stetson}) 
permits creation of the variable PSF, 
which was necessary for the
U-band images that showed fixed pattern residuals upon subtraction of the
non-variable PSF. In order to create a variable PSF I used 35 stars per
image.

Because of their slightly larger FWHM and
non-stellar PSF, globular clusters 
were easily detected on residual images created
by subtracting the objects fitted with a stellar PSF. This subtraction was
performed using the {\em allstar} task. Globular cluster
residuals were oversubtracted in the center and undersubtracted in the
wings (Fig.~\ref{residuals}). 
Images were visually inspected and the residuals in the U- and V-band
images compared. This was necessary to discard obvious galaxies from the
candidate list. 

After visual inspection of the residual
images, 50 and 81 objects in Field~1 and 2, respectively, were 
retained as candidates for globular clusters. Good globular cluster
candidates, the ones that have round ring-like residuals, are listed in
Tab.~\ref{gcdata}. Other extended objects for which the visual
distinction was not very clear, some of them being blended stars
or round galaxies, but some of which might be
globular clusters as well, are listed in Tab.~\ref{nogcdata}.
The Sextractor programme (Bertin \& Arnouts \cite{sextractor})
was used to measure precise coordinates, FWHM ($f_U$ and $f_V$ columns in
Tab.~\ref{gcdata} and ~\ref{nogcdata}) and ellipticities ($\epsilon_V$ and
$\epsilon_U$ columns) for these
objects.

I performed aperture photometry within IRAF for all the objects. 
Since the fields were quite crowded, stellar objects had to be
removed in order to obtain more precise photometry of globular clusters.
To do so, I first subtracted all the stars (this time excluding the
globular cluster candidates from the aperture photometry list) 
from the original
images using the {\em allstar} task. Aperture photometry of globular cluster
 candidates
was then measured through circular 
apertures of 16, 20 and 25 pixel radii. In some cases a bright saturated star
or another globular cluster
 candidate was found near the cluster, or the latter was located near
the edge of the image and thus only a 3 pix aperture magnitude was used.
The 3 pix magnitudes were later extrapolated to 25 pix magnitude values
using an aperture correction calculated as a median
correction for more isolated globular clusters. 
The error in magnitude for these
clusters is larger and hard to quantify, since the aperture correction
depends on the concentration and compactness of the cluster. 
The positions and magnitudes of globular clusters 
are presented in Tab.~\ref{gcdata} and
Tab.~\ref{nogcdata}. 
The clusters with
magnitudes measured through a smaller aperture and later corrected are
flagged with ``b'' (near the bright star or another globular cluster) 
or ``e'' (near the edge of the image).

\subsection{Photometric calibration}

\begin{figure}
    \resizebox{\hsize}{!}{\includegraphics[angle=270]{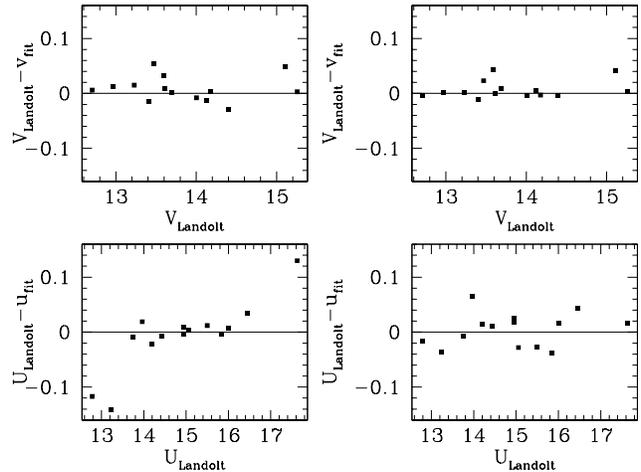}}
    \caption[]{Photometric calibration of Landolt (\cite{landolt}) stars during
the nights of observations in U and V filters. Left panels:
the scatter for the calibration without the color term;
right panels: the scatter for the calibration with the color term.
}
    \label{calib}  
\end{figure}

\begin{figure}
    \resizebox{\hsize}{!}{\includegraphics[angle=270]{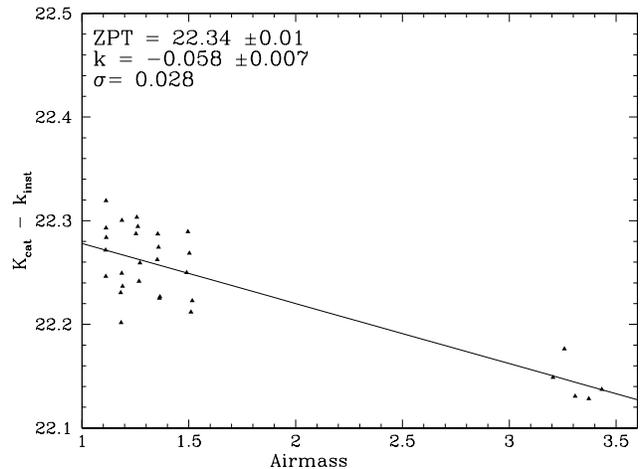}}
    \caption[]{Photometric calibration of
Persson {\it et al.}\ (\cite{persson}) stars
in K$_s$ filter.
}
    \label{Kcalib}  
\end{figure}

For the photometric calibration of the VLT images, a 
total of 14 stars in 4 different Landolt
(\cite{landolt})
fields, spanning the color range $-1.321<(U-V)< 4.162$, 
were observed during the two nights. 
The following transformations were
calculated:
\begin{eqnarray}
u_{inst} & = & U - 24.262(\pm0.089 ) + 0.379 (\pm0.068 )*X \nonumber \\
         &   & -0.042(\pm0.007)*(U-V) 
\label{Utran}
\end{eqnarray}

\begin{equation}
v_{inst} = V - 27.348 (\pm 0.042) + 0.213 (\pm 0.034)*X
\label{Vtran}
\end{equation}
where X is the mean airmass of the observations, $u_{inst}$ and $v_{inst}$ are the
instrumental magnitudes and U and V magnitudes from the Landolt
(\cite{landolt}) catalogue.
The one sigma scatter around the mean was 0.031 mag for the 
U band and 0.023 mag
for the V band (Fig.~\ref{calib} left panel). 
Adding the color term (U$-$V) in the transformations slightly 
reduces the scatter in the V-band to 0.017 mag (Fig.~\ref{calib} 
right panel). However, the calibration equation without the color term for
the V-band was prefered, since the data in that filter go much deeper and U
magnitudes for some objects could not be measured accurately enough.

SOFI images were calibrated using 6 standards from the
Persson {\it et al.}\ 
(\cite{persson}) list observed during the night and spanning 
 an airmass range from 1.1 to 3.4. Each standard star 
was observed at 5 different positions on the IR-array. In this way total of 
30 independent measurements was obtained. 
The least square fit (Fig.~\ref{Kcalib}) yielded the following 
calibration equation:
\begin{equation}
k_{s,inst} = K_s - 22.34 (\pm 0.01) + 0.058 (\pm 0.007) * X
\label{Ktran}
\end{equation}

\subsection{Completeness}
\label{completenesstest}

I made completeness tests for detection of stars on the U- and V-band
images. Using the IRAF task {\it addstar} within the DAOPHOT package I
added 400 stars each time to each of the U- and V-band images, 
 all with the
same magnitude and uniformly distributed over the whole field, and
re-computed the photometry. The completeness and the accuracy of 
photometry were measured at 0.5 mag bins.
The 90\% completeness for stellar sources
is reached at magnitude 25 in the 
V-band and 23.5 in U-band. The completeness drops to 50\% at
V$\sim26$ and U$\sim24.5$. More detailed results of completeness tests 
will be presented in the forthcoming paper where the stellar photometry 
of NGC~5128 is analyzed (Rejkuba {\it et al.}\
\cite{rejkuba2}).

Globular clusters have larger PSFs than stars, their selection is based
mainly on the recognition of a ``globular-like'' residual on a subtracted
image and thus their identification as such is more subjective. 
In order to check the completeness level of
globular cluster detections, I made similar simulations 
as for the stars.
The fainter globular clusters, that are of interest in simulation, are
not resolved and thus I just treat them as stars. To add a correct, typical
globular cluster PSF, the best cluster 
candidates' PSFs were used to create simulated images.
In order to investigate the dependence of detection on core radius,
simulations with different typical PSFs were made.
Two sets of simulations were done using the PSFs of clusters
with small core radii, f1.GC-17 and f2.GC-4,
to add objects, while in the other two sets much less
concentrated clusters f1.GC-10 and f2.GC-36 were used. 
At each pass 100 clusters were added, 
distributed uniformly all with the same magnitude.
The completeness of the globular cluster detection was measured for
each half magnitude. 

The detection depends quite significantly on core radius in the sense 
that more small $r_c$ clusters are lost at faint end. 
The incompleteness magnitude is defined as magnitude where more than 50\%
of the added clusters are not detected as such (they are actually all
detected, but their residuals are not recognized as globular cluster-like,
because they are lost in noise). It sets in at V$\sim$22.7 mag for clusters
like f1.GC-17 and f2.GC-4, but most of the larger clusters, like f1.GC-10
and f2.GC-36 can still be detected at 90\% level at magnitude V=23. The
incompleteness for clusters with large $r_c$ sets in at magnitude V=23.5. In
the simulations clusters could be detected with magnitudes as faint as V=24.
However, at those magnitude limits it is almost impossible to distinguish
globular cluster residuals from galaxies.

From the simulations I conclude that 
clusters fainter than $V\sim23$ mag would probably 
not be detected. MW clusters
corresponding to these magnitudes are Pal~1, 
Pal~13, Terzan~1, E~3 and AM~4.
However, at magnitude $V=22.0$ ($\sim2$ magnitudes past the GCLF peak)
more than 90\% of the clusters are detected.

\subsection{Comparison with published data}
\label{compliterature}

\begin{table}
  \caption[]{Comparison with the data from literature.}
    \label{literature.comp}
      \begin{tabular}{lllllll}
        \hline \hline
ID1 & ID2 & $B_{pg}$   & $V_{pg}$ & $V_{Wash}$ & $V_{WFI}$ & (U$-$V) \\
        \hline \hline
f1.GC-13 & C36 & 19.04 & 18.24   & 18.27 & 18.39 & 0.91 \\
f1.GC-16 & C37 & 19.24 & 18.17   & 18.46 & 18.47 & 1.33 \\
f1.GC-23 & C38 & 19.00 & \nodata & \nodata & 18.40 & 1.16 \\
f2.GC-26 & C15 & 19.76 & 18.62   & \nodata & 18.58 & 1.40 \\
f2.GC-81 & C12 & 18.82 & 17.74   & 17.79 & 18.01 & 1.47 \\
        \hline
      \end{tabular}
\end{table}

The most recent studies of the GCS in NGC~5128 (Holland {\it et al.}
\cite{holland}, Alonso \& Minniti \cite{am97}) have investigated the central
regions of the galaxy away from the fields studied here. 
The only photometry of globular
clusters found in common with the data presented here 
is the photographic plate
measurements of five clusters in
Hesser {\it et al.} (\cite{hetal86}, \cite{hetal84}) 
and the Washington photometry of
the three of these 5 clusters in 
Harris {\it et al.} (\cite{hetal92}). The clusters in common are C36, C37,
C38, C12 and C15 (numbering from Hesser {\it et al.} \cite{hetal84},
\cite{hetal86}). While in the 
Hesser {\it et al.} papers only the $\langle B
\rangle$ magnitude is
given, Harris {\it et al.} (\cite{hetal92}) report also photographic 
V-band magnitudes 
as well as Washington photometry for C12, C36 and C37. In order to compare
the Washington photometry with Bessell V magnitudes, I used the
transformation equation (Harris {\it et al.} \cite{hetal92}): $V = T_1 +
0.66 (M-T_1)$. 

The clusters in common are among the brightest ones and in this study
their magnitudes were
measured on 2.2m+WFI images. The comparison between the data from the 
literature and my measurements is given in Tab.~\ref{literature.comp}. In
columns 1 and 2 the identification of the objects in common is given,
$B_{pg}$ magnitudes are from Hesser {\it et al.} (\cite{hetal84},
\cite{hetal86}), $V_{pg}$ and $V_{Wash}$ are photographic magnitudes and the
V-band brightness derived from Washington photometry. The last two
columns present my measurements. 
There is a good agreement between the V-band magnitude derived from the
Washington photometry and my measurements. The largest difference is for 
cluster C12 (f2.GC-81), which I measured 0.2 magnitudes fainter, 
probably due to the vicinity of a bad column on the WFI chip. 
 
%
%

\begin{table*}
  \caption[]{Coordinates, photometry, FWHM ($f_V$ and $f_U$), ellipticities
($\epsilon_V$ and $\epsilon_U$) and projected galactocentric distance 
of globular cluster candidates. }
    \label{gcdata}
\begin{tabular}{lllllllllllllll}
\hline
\hline
 name  &alpha       &delta       &V     &$\sigma_V$ &U     &$\sigma_U$ &$K_s$ &
$\sigma_Ks$ &
$f_V$ &
$f_U$& $\epsilon_V$ & $\epsilon_U$ & R    & c\\
        &(2000)      &(2000)     &mag   &mag        &mag   &mag        &mag&
mag        &
pix      &
pix     &              &              &kpc   &        \\
\hline
\hline
f1.GC-1  &13:26:26.32 &-42:48:55.8 &19.58 &0.00 &20.49 &0.04 &\nodata&
\nodata& 8.70     &
7.40     & 0.21       & 0.24           &17.07 &\\
f1.GC-2  &13:26:29.01 &-42:49:44.7 &20.77 &0.03 &23.26 &0.30
&\nodata&\nodata& 4.11     &
4.34     & 0.05       & 0.3            &16.79 &\\
f1.GC-3  &13:26:26.99 &-42:49:50.4 &22.41 &0.04 &24.5: &0.4
&\nodata&\nodata& 5.17     &
\nodata  & 0.11       & \nodata        &16.45 &u\\
f1.GC-4  &13:26:28.13 &-42:51:29.5 &21.02 &0.03 &24.17 &0.31
&\nodata&\nodata& 4.09     &
3.49:    & 0.04       & 0.4:           &15.41 &u,n\\
f1.GC-5  &13:26:33.52 &-42:51:00.6 &19.52 &0.05 &21.11 &0.03
&\nodata&\nodata& 3.32     &
3.46     & 0.05       & 0.04           &16.53 &b\\
f1.GC-6  &13:26:40.76 &-42:50:18.2 &21.10 &0.10 &24.1: &0.4
&\nodata&\nodata& 4.23     &
\nodata  & 0.11       & \nodata        &18.07 &u,b,n\\
f1.GC-7  &13:26:21.10 &-42:48:41.4 &19.59 &0.00 &21.32 &0.04
&\nodata&\nodata& 4.61     &
4.77     & 0.06       & 0.09           &16.62 & \\
f1.GC-8  &13:26:19.88 &-42:49:11.5 &20.27 &0.10 &19.44 &0.05
&\nodata&\nodata &3.80     &
3.84     & 0.15       & 0.27           &16.06 &b,n\\
f1.GC-9  &13:26:21.36 &-42:49:59.6 &20.79 &0.01 &21.69 &0.05
&\nodata&\nodata& 3.45     &
3.63     & 0.07       & 0.09           &15.60 & \\
f1.GC-10 &13:26:14.80 &-42:50:09.1 &20.80 &0.01 &21.74 &0.09
&\nodata&\nodata& 5.07     &
5.33     & 0.23       & 0.32           &14.67 & \\
f1.GC-11 &13:26:09.76 &-42:50:29.9 &20.30 &0.09 &21.47 &0.07
&\nodata&\nodata& 4.33     &
4.66     & 0.03       & 0.20           &13.80 & b\\
f1.GC-12 &13:26:08.77 &-42:51:28.5 &19.70 &0.05 &21.15 &0.07
&\nodata&\nodata& 4.30     &
4.86     & 0.16       & 0.14           &12.87 & \\
f1.GC-13 &13:26:07.80 &-42:52:00.3 &18.39 &0.10 &19.30 &0.01
&\nodata&\nodata& \nodata  &
3.45     & \nodata    & 0.19           &12.31 & v\\
f1.GC-14 &13:26:09.45 &-42:53:17.4 &19.93 &0.05 &20.85 &0.02
&\nodata&\nodata& 4.01     &
4.22     & 0.05       & 0.21           &11.52 & b\\
f1.GC-15 &13:26:08.96 &-42:53:42.4 &19.44 &0.01 &20.93 &0.06
&\nodata&\nodata& 4.49     &
5.32     & 0.09       & 0.08           &11.14 &  \\
f1.GC-16 &13:26:10.65 &-42:53:42.4 &18.47 &0.10 &19.80 &0.02
&\nodata&\nodata& \nodata  &
3.58     & \nodata    & 0.19           &11.37 & v\\
f1.GC-17 &13:26:22.03 &-42:53:45.0 &19.58 &0.01 &20.36 &0.03
&\nodata&\nodata& 3.42     &
3.49     & 0.08       & 0.15           &13.02 & \\
f1.GC-18 &13:26:22.04 &-42:54:25.9 &19.67 &0.01 &20.45 &0.04
&\nodata&\nodata& 3.44     &
3.58     & 0.07       & 0.15           &12.61 & \\
f1.GC-19 &13:26:05.27 &-42:54:58.1 &18.75 &0.10 &19.52 &0.01
&\nodata&\nodata& \nodata  &
3.29     & \nodata    & 0.12           &9.72  & v\\
f1.GC-20 &13:26:05.45 &-42:55:22.4 &21.51 &0.05 &23.33 &0.10
&\nodata&\nodata& 5.52     &
4.29:    & 0.06       & 0.5:           &9.47  & u,e\\
f1.GC-21 &13:26:10.60 &-42:55:00.0 &21.65 &0.03 &23.72 &0.28
&\nodata&\nodata& 4.16     &
3.89:    & 0.07       & 0.4:           &10.48 & u\\
f1.GC-22 &13:26:15.95 &-42:55:00.5 &18.09 &0.10 &18.95 &0.01
&\nodata&\nodata& \nodata  &
3.34     & \nodata    & 0.17           &11.30 & v\\
f1.GC-23 &13:26:23.81 &-42:54:00.5 &18.40 &0.10 &19.56 &0.01
&\nodata&\nodata& \nodata  &
3.32     & \nodata    & 0.17           &13.14 & v\\
f1.GC-24 &13:26:28.87 &-42:52:35.9 &19.33 &0.00 &20.17 &0.02
&\nodata&\nodata& 3.73     &
3.84     & 0.12       & 0.11           &14.8  &  \\
f1.GC-34 &13:26:09.76 &-42:53:16.8 &21.27 &0.10 &23.22 &0.30
&\nodata&\nodata& 4.45     &
5.63     & 0.09       & 0.32           &11.57 & b\\
f1.GC-38 &13:26:20.06 &-42:55:22.8 &21.84 &0.10 &23.54 &0.30
&\nodata&\nodata& 5.29     &
4.49:    & 0.13       & 0.46:          &11.75 & u,e\\
 f2.GC-1 &13:25:42.07 &-43:12:44.2 &23.10 &0.07 &24.9: &0.4
&\nodata&\nodata& 3.79     &
\nodata  & 0.03       & \nodata    &12.43 & u\\
 f2.GC-2 &13:25:41.92 &-43:10:40.7 &19.52 &0.10 &20.40 &0.03
&\nodata&\nodata& \nodata  &
3.61     & \nodata    & 0.14       &10.33 & v\\
 f2.GC-3 &13:25:40.82 &-43:08:15.1 &20.10 &0.10 &20.62 &0.03
&\nodata&\nodata& \nodata  &
3.56     & \nodata    &  0.13      &7.84  & v\\
 f2.GC-4 &13:25:40.46 &-43:07:16.9 &19.77 &0.01 &21.13 &0.06
&\nodata&\nodata& 2.81     &
3.59     & 0.05       & 0.08       &6.87  &  \\
 f2.GC-7 &13:25:39.04 &-43:12:56.7 &22.41 &0.05 &25.8: &0.4
&\nodata&\nodata& 3.45     &
\nodata  & 0.13       & \nodata    &12.53 & u,n\\
 f2.GC-8 &13:25:38.97 &-43:10:40.9 &21.55 &0.02 &23.27 &0.14 &
\nodata&\nodata& 3.62     &
5.95     & 0.07       & 0.08       &10.20 &  \\
 f2.GC-9 &13:25:38.10 &-43:13:01.2 &20.59 &0.01 &21.66 &0.07
&\nodata&\nodata& 3.07     &
4.00     & 0.06       & 0.13       &12.57 & \\
f2.GC-10 &13:25:37.58 &-43:12:32.3 &23.16 &0.10 &23.90 &0.20
&\nodata&\nodata& 3.15     &
3.68     & 0.15       & 0.38       &12.06 & u\\
f2.GC-11 &13:25:36.65 &-43:12:49.9 &22.30 &0.04 &26.0: &0.4
&\nodata&\nodata& 3.49     &
\nodata  & 0.05       & \nodata    &12.34 & u,n\\
f2.GC-13 &13:25:35.84 &-43:07:27.2 &20.96 &0.02 &22.60 &0.10
&\nodata&\nodata& 2.66     &
3.32     & 0.07       & 0.13       & 6.77 & \\
f2.GC-14 &13:25:35.82 &-43:07:15.3 &22.08 &0.02 &22.70 &0.10
&\nodata&\nodata& 2.85     &
4.17     & 0.06       & 0.16       & 6.57 & \\
f2.GC-15 &13:25:35.55 &-43:08:36.1 &20.83 &0.01 &21.99 &0.07 &18.29 & 0.10&
 2.88     &
3.72     & 0.06       & 0.05       & 7.94 & \\
f2.GC-17 &13:25:34.01 &-43:10:44.6 &20.48 &0.10 &21.43 &0.05 &17.60 & 0.10&
 2.73     &
3.50     & 0.08       & 0.05       &10.10 &  b\\
f2.GC-18 &13:25:33.58 &-43:07:18.5 &21.44 &0.04 &22.41 &0.09 &17.96 & 0.19&
 2.85     &
4.02     & 0.05       & 0.12       & 6.54 &  \\
f2.GC-20 &13:25:33.07 &-43:07:00.6 &21.26 &0.04 &22.84 &0.04 &17.68 & 0.09&
 3.18     &
4.62     & 0.07       & 0.18       & 6.21 & b \\
f2.GC-21 &13:25:33.03 &-43:09:20.8 &20.84 &0.01 &21.54 &0.05 &18.18 & 0.11&
 2.86     &
3.70     & 0.04       & 0.09       & 8.63 & \\
f2.GC-22 &13:25:32.76 &-43:11:51.1 &21.13 &0.04 &25.4: &0.5
&\nodata&\nodata& 3.26     &
\nodata  & 0.05       & \nodata    &11.23 & u,n\\
f2.GC-23 &13:25:32.71 &-43:07:01.6 &18.80 &0.10 &20.03 &0.02 & 15.60& 0.03&
 \nodata  &
3.50     & \nodata    & 0.16       & 6.22 & v,b\\
f2.GC-24 &13:25:32.25 &-43:07:16.5 &20.12 &0.01 &20.95 &0.04 & 17.71& 0.08&
2.83     &
3.56     & 0.05       & 0.08       & 6.46 &  \\
f2.GC-26 &13:25:30.35 &-43:11:48.4 &18.58 &0.10 &19.98 &0.02 &
\nodata&\nodata&\nodata  &
3.88     & \nodata    & 0.19       &11.15 & v\\
f2.GC-28 &13:25:30.10 &-43:06:54.0 &21.49 &0.03 &22.56 &0.16 & 17.05& 0.08&
3.49     &
4.35     & 0.13       & 0.08       & 6.02 & \\
f2.GC-29 &13:25:29.68 &-43:11:41.7 &19.76 &0.10 &20.62 &0.04 &
\nodata&\nodata&\nodata  &
3.61     & \nodata    & 0.15       &11.03 & v\\
f2.GC-31 &13:25:29.38 &-43:07:41.1 &20.66 &0.02 &21.47 &0.08 & 18.24& 0.12&
2.97     &
3.74     & 0.09       & 0.13       & 6.84 & \\
f2.GC-34 &13:25:28.06 &-43:06:46.3 &22.40 &0.09 &23.12 &0.17 &
\nodata&\nodata&3.17     &
3.68     & 0.02       & 0.15       & 5.87 & \\
f2.GC-35 &13:25:26.75 &-43:09:39.7 &19.54 &0.00 &20.44 &0.03 & 17.00& 0.05&
3.40     &
4.01     & 0.10       & 0.12       & 8.90 & \\
f2.GC-36 &13:25:26.68 &-43:08:52.7 &19.72 &0.01 &20.57 &0.03 & 17.12& 0.06&
4.50     &
5.13     & 0.09       & 0.08       & 8.08 & \\
f2.GC-40 &13:25:24.32 &-43:07:58.5 &19.59 &0.10 &20.58 &0.04 & 16.72& 0.04&
 \nodata &
3.75     & \nodata    & 0.16       & 7.15 & v\\
f2.GC-41 &13:25:22.61 &-43:07:37.6 &21.43 &0.04 &23.29 &0.19 & 17.65& 0.09&
3.85     &
5.36     & 0.08       & 0.16       & 6.83 &  \\
f2.GC-43 &13:25:21.26 &-43:10:03.0 &21.41 &0.02 &21.78 &0.10 & 17.62& 0.12&
4.91     &
5.48     & 0.12       & 0.13       & 9.38 &  \\
f2.GC-46 &13:25:20.63 &-43:06:35.6 &20.27 &0.20 &21.50 &0.20 &
\nodata&\nodata&3.11     &
3.73     & 0.07       & 0.07       & 5.83 & e\\
f2.GC-47 &13:25:19.90 &-43:07:43.5 &20.22 &0.01 &21.78 &0.05 & 16.75& 0.04&
3.25     &
4.04     & 0.05       & 0.15       & 7.02 &  \\

\hline
\end{tabular}

\end{table*}

\addtocounter{table}{-1}
\begin{table*}
  \caption[]{continuation.}
\begin{tabular}{lllllllllllllll}
\hline
\hline
 name  &alpha       &delta       &V     &$\sigma_V$ &U     &$\sigma_U$ &
$K_s$ &
$\sigma_Ks$& $f_V$ &
$f_U$& $\epsilon_V$ & $\epsilon_U$ & R    & c\\
        &(2000)      &(2000)      &mag   &mag  &mag   &mag  & mag & mag &
pix      & pix     &              &              &kpc   &        \\
\hline
\hline

f2.GC-48 &13:25:19.42 &-43:11:51.0 &21.09 &0.02 &21.63 &0.09 &
\nodata&\nodata&3.99     &
4.89     & 0.21       & 0.13       &11.29 &  \\
f2.GC-49 &13:25:18.79 &-43:07:11.1 &22.21 &0.23 &23.59 &0.15 & 18.88& 0.28&
5.07     &
\nodata  & 0.16       & \nodata    & 6,52 & b,u\\
f2.GC-50 &13:25:18.79 &-43:10:53.7 &22.61 &0.08 &24.4: &0.3  & 18.39& 0.11&
3.31     &
\nodata  & 0.06       & \nodata    &10.32 &  u\\
f2.GC-52 &13:25:17.84 &-43:13:21.4 &21.94 &0.12 &22.44 &0.08 &
\nodata&\nodata&4.45     &
6.57     & 0.19       & 0.31       &12.90 & \\
f2.GC-53 &13:25:17.28 &-43:08:38.6 &19.87 &0.10 &20.69 &0.03 & 16.61& 0.08&
3.11     &
3.50     & 0.07       & 0.12       & 8.07 & b\\
f2.GC-54 &13:25:16.91 &-43:09:27.4 &19.48 &0.10 &20.41 &0.03 & 16.55& 0.05&
\nodata  &
3.67     & \nodata    & 0.10       & 8.91 & v \\
f2.GC-57 &13:25:15.19 &-43:08:38.8 &20.86 &0.02 &22.10 &0.05 & 17.58& 0.08&
2.86     &
3.72     & 0.07       & 0.14       & 8.18 &  \\
f2.GC-59 &13:25:13.82 &-43:07:32.3 &20.61 &0.02 &22.04 &0.11 & 17.43& 0.07&
3.09     &
4.01     & 0.05       & 0.08       & 7.17 & b \\
f2.GC-60 &13:25:12.90 &-43:06:39.2 &20.50 &0.05 &23.52 &0.16 &
\nodata&\nodata&2.51     &
4.30     & 0.04       & 0.01       & 6.39 & n \\
f2.GC-61 &13:25:12.85 &-43:07:58.9 &18.23 &0.10 &19.64 &0.01 & 15.12& 0.02&
\nodata  &
3.46     & \nodata    & 0.14       & 7.67 & v\\
f2.GC-63 &13:25:12.76 &-43:10:42.7 &20.75 &0.02 &22.39 &0.19 & 17.77& 0.08&
4.45     &
5.95     & 0.09       & 0.19       &10.38 &  \\
f2.GC-65 &13:25:10.39 &-43:08:11.1 &22.74 &0.09 &23.73 &0.25 &
\nodata&\nodata&3.00     &
5.23:    & 0.07       & 0.43:      & 8.04 &  u\\
f2.GC-66 &13:25:10.11 &-43:06:41.7 &21.05 &0.02 &23.43 &0.15 &
\nodata&\nodata&2.50     &
4.79:    & 0.01       & 0.20       & 6.68 & \\
f2.GC-69 &13:25:09.07 &-43:10:01.4 &19.87 &0.01 &21.23 &0.04 &
\nodata&\nodata&2.78     &
3.64     & 0.10       & 0.07       & 9.92 &  \\
f2.GC-70 &13:25:08.92 &-43:08:53.3 &20.30 &0.01 &21.33 &0.05 &
\nodata&\nodata&3.67     &
4.51     & 0.06       & 0.06       & 8.83 &  \\
f2.GC-71 &13:25:08.78 &-43:09:09.1 &19.42 &0.10 &20.43 &0.03
&\nodata&\nodata& \nodata   &
3.88     & \nodata    & 0.12       & 9.09 &  v\\
f2.GC-74 &13:25:07.48 &-43:12:28.6 &20.43 &0.01 &21.40 &0.07 &
\nodata&\nodata&2.89     &
4.18     & 0.12       & 0.19       &12.44 &  \\
f2.GC-75 &13:25:07.41 &-43:07:34.8 &21.83 &0.04 &23.36 &0.17 &
\nodata&\nodata&4.10     &
5.80     & 0.13       & 0.40       & 7.74 &  u\\
f2.GC-76 &13:25:07.31 &-43:08:29.3 &20.67 &0.01 &21.80 &0.07 &
\nodata&\nodata&3.78     &
4.28     & 0.02       & 0.15       & 8.58 &  \\
f2.GC-79 &13:25:05.88 &-43:12:02.2 &22.70 &0.07 &23.97 &0.37 &
\nodata&\nodata&3.19     &
5.25:    & 0.11       & 0.46:      &12.11 & u\\
f2.GC-81 &13:25:05.70 &-43:10:30.2 &18.01 &0.20 &19.48 &0.01 &
\nodata&\nodata&\nodata   &
3.83     & \nodata    & 0.11       &10.63 & v\\
\hline
\end{tabular}
\begin{tabular}{llll}
comment: &b &= &near bright (saturated) star \\
 &e &= &near edge of the field \\
 &u &= &marginal or no detection \\
 &v &= &object saturated in V; V mag from WFI data \\
 &n &= &object does not satisfy all the criteria for globular cluster
classification\\
\end{tabular}

\end{table*}

\begin{table*}
  \caption[]{Same as Tab.~\ref{gcdata}, but for other
extended objects, some of which might as well be globular clusters.}
    \label{nogcdata}
\begin{tabular}{lllllllllllllll}
\hline
\hline
 name  &alpha       &delta       &V     &$\sigma_V$ &U     &$\sigma_U$
&$K_s$ & $\sigma_Ks$& $f_V$ &
$f_U$& $\epsilon_V$ & $\epsilon_U$ & R    & c\\
        &(2000)      &(2000)      &mag   &mag  &mag   &mag  & mag & mag &
pix      &
pix     &              &              &kpc   &        \\
\hline
\hline
f1.GC-25 &13:26:28.27 &-42:50:03.9 &20.43 &0.10 &19.68 &0.10 &
\nodata&\nodata&4.65     &
4.42     & 0.29       & 0.32           &16.45 &  \\
f1.GC-26 &13:26:25.48 &-42:51:07.5 &21.15 &0.11 &21.23 &0.04 &
\nodata&\nodata&7.63     &
9.06     & 0.39       & 0.33           &15.3  &  \\
f1.GC-27 &13:26:30.45 &-42:52:10.8 &18.33 &0.10 &19.19 &0.01 &
\nodata&\nodata&\nodata  &
3.62     & \nodata    & 0.31           &15.3  & v \\
f1.GC-28 &13:26:14.46 &-42:49:34.1 &20.20 &0.05 &24.13 &0.30 &
\nodata&\nodata&4.85     &
\nodata  & 0.18       & \nodata        &15.12 & u \\
f1.GC-29 &13:26:07.23 &-42:51:38.1 &17.09 &0.10 &18.92 &0.01 &
\nodata&\nodata&\nodata  &
2.84     & \nodata    & 0.11           &12.55 & v \\
f1.GC-30 &13:26:16.36 &-42:54:28.5 &21.37 &0.02 &21.84 &0.10 &
\nodata&\nodata&6.05     &
7.33     & 0.38       & 0.55           &11.69 &  \\
f1.GC-31 &13:26:17.40 &-42:54:14.5 &19.24 &0.10 &19.76 &0.02 &
\nodata&\nodata& \nodata  &
2.99     & 0.10       & \nodata        &12.00 & v \\
f1.GC-32 &13:26:06.37 &-42:53:54.4 &20.74 &0.04 &21.20 &0.03 &
\nodata&\nodata&2.83     &
3.26     & 0.08       & 0.23           &10.64 &   \\
f1.GC-33 &13:26:04.74 &-42:54:00.6 &22.90 &0.10 &22.91 &0.20 &
\nodata&\nodata&4.53     &
5.51     & 0.14       & 0.38           &10.35 & u,e \\
f1.GC-35 &13:26:07.81 &-42:55:05.6 &19.95 &0.00 &21.19 &0.04 &
\nodata&\nodata&2.80     &
3.15     & 0.08       & 0.18           &10.00 &  \\
f1.GC-36 &13:26:11.89 &-42:54:51.8 &16.95 &0.10 &19.35 &0.01
&\nodata&\nodata& \nodata  &
3.10     & \nodata    & 0.21           &10.76 & v \\
f1.GC-37 &13:26:14.14 &-42:54:37.6 &20.91 &0.02 &21.09 &0.04 &
\nodata&\nodata&5.41     &
6.56     & 0.25       & 0.32           &11.26 &  \\
f1.GC-39 &13:26:26.66 &-42:55:19.3 &17.01 &0.10 &19.49 &0.01 &
\nodata&\nodata&\nodata  &
2.93     & \nodata    & 0.21           &12.88 & v \\
f1.GC-40 &13:26:09.63 &-42:54:26.2 &22.32 &0.05 &23.22 &0.23 &
\nodata&\nodata&4.15     &
4.11     & 0.11       & 0.09           &10.71 &  \\
f1.GC-41 &13:26:07.05 &-42:52:37.5 &17.14 &0.10 &18.07 &0.01 &
\nodata&\nodata&\nodata  &
2.98     & \nodata    & 0.17           &11.72 & v \\
f1.GC-42 &13:26:05.18 &-42:52:34.0 &16.59 &0.10 &17.94 &0.00 &
\nodata&\nodata&\nodata  &
3.10     & \nodata    & 0.18           &11.55 & v \\
f1.GC-43 &13:26:06.78 &-42:52:43.2 &17.62 &0.10 &18.30 &0.01 &
\nodata&\nodata&\nodata  &
2.97     & \nodata    & 0.15           &11.61 & v \\
f1.GC-44 &13:26:11.93 &-42:53:02.9 &18.65 &0.10 &19.25 &0.01 &
\nodata&\nodata&\nodata  &
2.94     & \nodata    & 0.14           &12.03 & v \\
f1.GC-45 &13:26:07.65 &-42:54:28.3 &17.76 &0.10 &19.16 &0.01 &
\nodata&\nodata&\nodata  &
3.04     & \nodata    & 0.20           &10.41 & v \\
f1.GC-46 &13:26:07.22 &-42:54:20.7 &17.43 &0.10 &18.24 &0.00 &
\nodata&\nodata&\nodata  &
3.16     & \nodata    & 0.22           &10.43 & v \\
f1.GC-47 &13:26:06.43 &-42:54:21.3 &18.88 &0.10 &20.26 &0.02 &
\nodata&\nodata&\nodata  &
3.10     & \nodata    & 0.22           &10.32 & v \\
f1.GC-48 &13:26:06.19 &-42:54:30.3 &18.54 &0.10 &19.25 &0.01 &
\nodata&\nodata&\nodata  &
3.08     & \nodata    & 0.20           &10.18 & v \\
f1.GC-49 &13:26:10.13 &-42:54:56.4 &19.89 &0.10 &21.16 &0.05 &
\nodata&\nodata&\nodata  &
3.27     & \nodata    & 0.22           &10.45 & v \\
f1.GC-50 &13:26:14.97 &-42:55:21.4 &17.57 &0.10 &18.26 &0.15 &
\nodata&\nodata&\nodata  &
3.08     & \nodata    & 0.23           &10.94 & v \\
 f2.GC-5 &13:25:40.13 &-43:10:13.3 &21.96 &0.03 &26.3: &0.5  &
\nodata&\nodata&4.12     &
\nodata  & 0.23       & \nodata    &9.78  & u \\
 f2.GC-6 &13:25:39.11 &-43:12:11.6 &21.34 &0.02 &23.11 &0.11 &
\nodata&\nodata&2.40     &
3.78     & 0.04       & 0.14       &11.76 & \\
f2.GC-12 &13:25:36.39 &-43:08:02.7 &21.47 &0.07 &22.35 &0.16 &
\nodata&\nodata&4.98     &
5.32     & 0.29       & 0.32       & 7.40 & \\
f2.GC-16 &13:25:34.40 &-43:12:04.2 &23.58 &0.13 &23.58 &0.21 &
\nodata&\nodata&3.95     &
5.24     & 0.16       & 0.46       &11.49 & u\\
f2.GC-19 &13:25:33.25 &-43:10:19.6 &22.67 &0.06 &24.6: &0.5  & 17.75 & 0.11&
4.01     &
\nodata  & 0.17       & \nodata    & 9.65 & u\\
f2.GC-25 &13:25:31.09 &-43:10:27.6 &21.13 &0.02 &24.34 &0.24 & 17.00 & 0.10&
3.54     &
4.01     & 0.10       & 0.09       & 9.75 & u\\
f2.GC-27 &13:25:30.30 &-43:11:26.6 &21.90 &0.03 &24.04 &0.30 &
\nodata&\nodata&4.84     &
\nodata  & 0.09       & \nodata    &10.77 & u\\
f2.GC-30 &13:25:29.42 &-43:09:21.9 &22.36 &0.06 &24.60 &0.26 & 20.03 & 0.47
&3.58     &
\nodata  & 0.03       & \nodata    & 8.59 & u\\
f2.GC-32 &13:25:29.00 &-43:12:50.9 &22.44 &0.05 &23.67 &0.19 &
\nodata&\nodata&6.74     &
3.40:    & 0.43       & 0.39       &12.23 & \\
f2.GC-33 &13:25:28.26 &-43:06:40.0 &22.31 &0.06 &24.71 &0.27 &
\nodata&\nodata&4.26     &
\nodata  & 0.04       & \nodata    & 5.76 & u\\
f2.GC-37 &13:25:24.77 &-43:06:54.8 &22.36 &0.09 &24.07 &0.26 & 17.84 & 0.18&
4.36     & 
\nodata  & 0.17       & \nodata    & 6.04 & u\\
f2.GC-38 &13:25:24.66 &-43:11:50.6 &21.95 &0.03 &23.48 &0.38 &
\nodata&\nodata&4.06     &
6.91:    & 0.10       & 0.49       &11.19 & u\\
f2.GC-39 &13:25:24.61 &-43:08:15.2 &22.97 &0.11 &26.1: &0.6  & 16.90 & 0.07&
4.16     &
\nodata  & 0.26       & \nodata    & 7.44 & u\\
f2.GC-42 &13:25:21.46 &-43:11:43.8 &20.80 &0.20 &21.80 &0.08 &
\nodata&\nodata&6.46     &
7.17     & 0.42       & 0.45       &11.12 & \\
f2.GC-44 &13:25:21.24 &-43:06:57.0 &20.70 &0.01 &24.66 &0.32 & 16.92 & 0.08&
2.42      &
4.31     & 0.03       & 0.21       & 6.17 & \\
f2.GC-45 &13:25:21.03 &-43:08:31.9 &21.83 &0.04 &21.98 &0.07 & 18.14 & 0.08&
5.72     &
5.57     & 0.29       & 0.25       & 7.81 & \\
f2.GC-51 &13:25:18.00 &-43:06:57.6 &22.09 &0.05 &23.07 &0.11 & 17.86 & 0.10&
4.23     &
5.19     & 0.39       & 0.31       & 6.33 & \\
f2.GC-55 &13:25:16.76 &-43:13:13.5 &19.38 &0.00 &21.16 &0.05 &
\nodata&\nodata&3.65     &
4.98     & 0.08       & 0.26       &12.79 & \\
f2.GC-56 &13:25:16.72 &-43:13:18.4 &20.63 &0.20 &22.19 &0.10 &
\nodata&\nodata&3.98     &
5.62     & 0.26       & 0.14       &12.88 & \\
f2.GC-58 &13:25:14.84 &-43:11:00.3 &22.81 &0.09 &22.95 &0.16 & 18.22 & 0.10&
4.58     &
6.75     & 0.33       & 0.41       &10.58 & \\
f2.GC-62 &13:25:12.81 &-43:12:00.7 &21.51 &0.05 &25.8: &0.5: &
\nodata&\nodata&5.55     &
\nodata  & 0.35       & \nodata    &11.70 & u\\
f2.GC-64 &13:25:12.17 &-43:08:48.7 &21.31 &0.10 &23.43 &0.20 & 17.30 & 0.09&
6.31     &
\nodata  & 0.15       & \nodata    & 8.52 & u\\
f2.GC-67 &13:25:09.32 &-43:09:58.4 &22.88 &0.10 &23.99 &0.12 &
\nodata&\nodata&5.15     &
\nodata  & 0.15       & \nodata    & 9.85 & u\\
f2.GC-68 &13:25:09.06 &-43:13:18.4 &20.44 &0.01 &22.00 &0.09 &
\nodata&\nodata& 7.25     &
8.77     & 0.46       & 0.27       &13.19 & \\
f2.GC-72 &13:25:08.11 &-43:10:37.4 &21.00 &0.03 &21.51 &0.07 &
\nodata&\nodata&6.21     &
6.69     & 0.36       & 0.19       &10.57 & \\
f2.GC-73 &13:25:08.03 &-43:09:51.8 &22.99 &0.12 &23.88 &0.17 &
\nodata&\nodata&5.32     &
\nodata  & 0.16       & \nodata    & 9.84 & u\\
f2.GC-77 &13:25:06.66 &-43:08:57.9 &21.16 &0.03 &23.41 &0.19 &
\nodata&\nodata&5.73     &
4.25     & 0.18       & 0.29       & 9.09 & u\\
f2.GC-78 &13:25:06.11 &-43:06:51.9 &22.06 &0.05 &24.4: &0.5  &
\nodata&\nodata&3.39     &
\nodata  & 0.08       & \nodata    & 7.23 & u\\
f2.GC-80 &13:25:05.72 &-43:06:50.9 &20.96 &0.02 &23.91 &0.40 &
\nodata&\nodata&2.52     &
4.51     & 0.08       & 0.18       & 7.26 &  \\
\hline
\end{tabular}
\begin{tabular}{llll}
comment: &b &= &near bright (saturated) star \\
 &e &= &near edge of the field \\
 &u &= &marginal or no detection \\
 &v &= &object saturated in V; V mag is from WFI data \\
\end{tabular}

\end{table*}

\subsection{Selection of globular clusters}
\label{selection}

\begin{figure}
    \resizebox{\hsize}{!}{\includegraphics[angle=270]{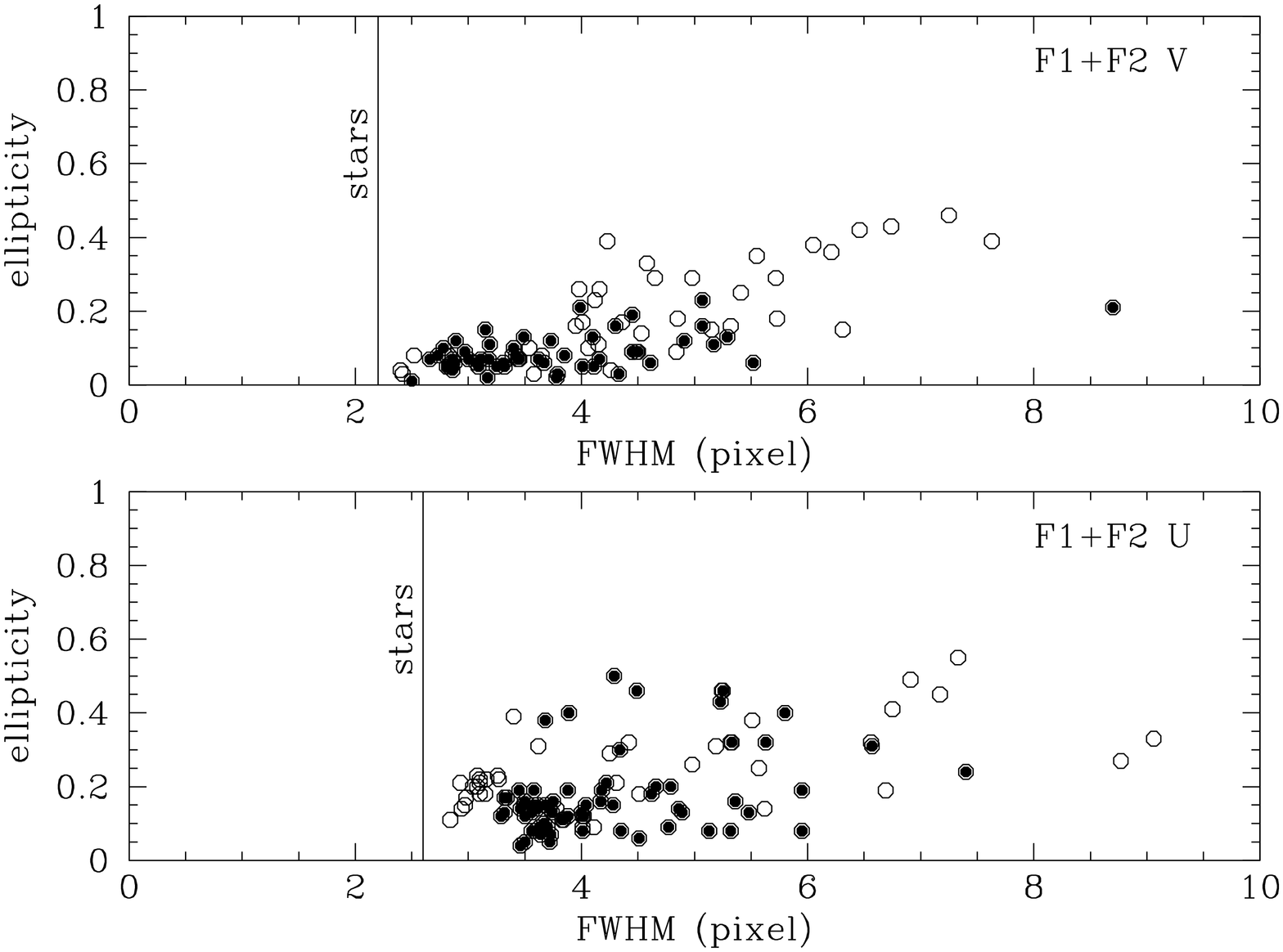}}
    \caption[]{FWHM {\it vs.}\ ellipticity: objects retained as good globular
cluster candidates after the visual inspection of the residual images
of Field~1 and 2 are
plotted with filled circles. Empty circles are used for other extended
objects.}
     \label{FWHMell}                                                    
\end{figure}

Typical Galactic globular clusters
 have mean King core radii $r_c=2.3\pm3.6$pc, mean
tidal radii $r_t=45.1 \pm 32.4$pc and mean ellipticities $\epsilon = 0.07 \pm
0.01$ (June 1999 update of Harris \cite{harris96} catalogue; 
White \& Shawl \cite{rwhite}). 

Holland {\it et al.}\ (\cite{holland}) used HST WFPC2 data to find globular
clusters in the central parts of NGC~5128 and measured their
structural parameters. They did not find any 
significant difference between the
MW and NGC~5128 globular clusters'
 King core radii ($r_c$), tidal radii ($r_t$) 
and half-mass radii ($r_h$). They also found no trend 
in $r_c$, $r_t$ or $r_h$ with galactocentric distance, 
contrary to what one may expect due to stronger tidal forces closer to 
the galactic center.
However, they measured higher mean ellipticities for their
globular cluster candidates with respect to the MW 
globular clusters. Their samples
are restricted only to the brighter clusters in NGC~5128, 
corresponding to the absolute magnitude $M_V=-6.4$. For comparison, the
present sample reaches almost 2 magnitudes deeper.

I measured the FWHM and ellipticity for all the objects in
Tab.\ ~\ref{gcdata} and ~\ref{nogcdata} using the Sextractor programme
(Bertin \& Arnouts \cite{sextractor}). U-band images had variable PSF
across the frame and the objects were, in some places, elongated, thus
showing higher ellipticity than in the V-band images.

The FWHM {\it vs.}\ ellipticity for all the objects in Fields~1 and 2 
is plotted in 
Fig.~\ref{FWHMell}. The upper panel shows the results obtained from the
V-band, the lower panel the ones obtained from
the U-band. The vertical line indicates the mean FWHM for stars. 
The objects from Tab.~\ref{nogcdata}, some of which might 
as well be blended stars or
galaxies are plotted as open symbols, while the filled dots were used for
objects from Tab.~\ref{gcdata} (i.e.\ good candidates). Objects
that were slightly saturated on the V-band VLT images are shown only 
in the
U-band diagram (lower panel). Due to the poorer 
seeing conditions during the
observations with the 2.2m telescope, their profiles are more 
similar to the stellar profiles on the WFI images.

In order
to classify an object as a globular cluster all of
the following criteria were imposed:
\begin{enumerate}
\item[1.] non-stellar PSF that leaves ring-like residuals on subtracted
images (Fig.~\ref{residuals});
\item[2.] FWHM$>$mean stellar FWHM and ellipticity(V)$<$0.4 (Fig.~\ref{FWHMell})
\item[3.] $0<(U-V)_0 < 2.5$ (Fig.~\ref{CMDgc}) 
\end{enumerate}
The selection of objects with ellipticity measured on the V-band images
smaller than 0.4 excludes galaxies from the sample. 
The color range in the last criterion corresponds to the observed range 
of the (U$-$V)$_0$ colors for the MW and M31 globular clusters. This range
does not include the reddest M31 clusters (Fig.~\ref{CMDgc}) because they
were corrected only for the foreground average extinction towards M31 and
thus might still be reddened. The cutoff at (U$-$V)$_0=2.5$ takes into
account the bulk of M31 clusters around $V_0\sim21$. 
The objects that did not 
pass all of the above criteria were flagged with ``n''
in Tab.~\ref{gcdata}.

\section{Color-Magnitude and Color-Color Diagrams}
\label{CMD}

\begin{figure}
    \resizebox{\hsize}{!}{\includegraphics{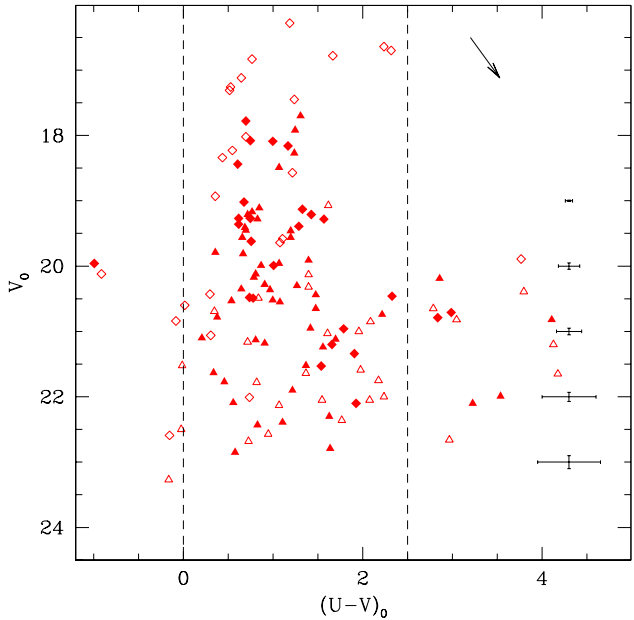}}
    \caption[]{Color-Magnitude Diagram for all objects found in Field~1
(diamonds)
and in Field~2 (triangles). Objects retained as good globular cluster 
candidates
after the visual inspection of images are plotted as filled
symbols and the rest as open ones.}
    \label{CMDgc.CenA}   
\end{figure}

The UV color-magnitude diagram (CMD) for all the objects
is presented in Fig.~\ref{CMDgc.CenA}. 
As before, the objects retained as good candidates after
visual inspection of the images are plotted as filled symbols and the rest
as open symbols. The objects from Field~1 are plotted as diamonds, the
ones from Field~2 as triangles.
Vertical lines indicate the color range within which the
MW and majority of M31 globular clusters are located.
Note that several open symbols are found in the same
area of the CMD as globular clusters. Spectroscopic
observations are necessary to assess the real nature of these objects.

The object at (U$-$V)$_0=-0.99$ and $V_0=19.96$ (f1.GC-8) 
is located very close to the
area where a large number of bright young blue stars were found (Rejkuba
{\it et al.}\ \cite{rejkuba1}). It might be a young object, precursor of
a globular cluster, similar to the ones found in other merging galaxies
(e.g.\ Whitmore {\it et al.}\ \cite{whitmore}), as well as to some clusters
found in the central regions of NGC~5128 (Holland {\it et al.}\
\cite{holland}).

There is a number of objects brighter than $V<17.5$ 
that were all identified in Field 1. Probably these are bright stars blended
with some faint companions. They all appear saturated in the 
V-band VLT images and 
were found in the lower right corner (south-west) of Field~1 on U-band
images. Their FWHM, that was larger than stellar, is
probably an artifact due to the distortion of the 
PSF in that part of the chip.

\begin{figure}
    \resizebox{\hsize}{!}{\includegraphics{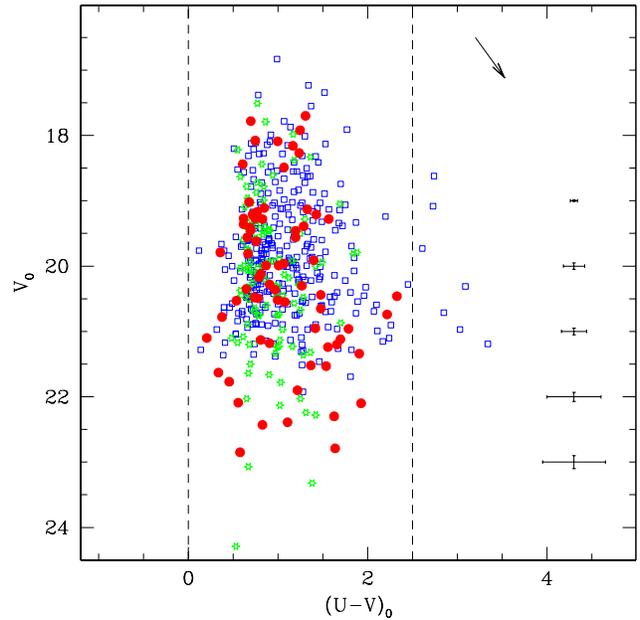}}
    \caption[]{Color-Magnitude Diagram for globular clusters
in NGC~5128 (filled circles)
compared to the MW (stars) and M31 globular clusters (open squares).}
    \label{CMDgc}
\end{figure}

Figure~\ref{CMDgc} shows the CMD for all NGC~5128 globular clusters 
(filled circles) that satisfy all the selection criteria. This CMD is 
compared with the ones of
the MW (stars) and M31 (squares) globular clusters. Globular clusters
populate a similar region of the CMD in the three galaxies. Data for the MW
clusters were taken from the web page of W.\ Harris (\cite{harris96}). M31
clusters are from the catalogue of Barmby {\it et al.}\ (\cite{barmby}).
I did not take into account globular clusters 
with colors bluer than B$-$V$=0.55$, as they
might be younger objects (Barmby {\it et al.}\ \cite{barmby}). 
The magnitudes were corrected for the mean reddening towards M31
(E(B$-$V$)=0.08$). For the globular cluster candidates in NGC~5128 a mean
foreground reddening of E(B$-$V$)=0.1$ (Burstein \& Heiles \cite{BH}; 
Schlegel {\it et al.}\ \cite{schlegel}) was used. 
The reddening vector plotted in the
upper right corner corresponds to E(B$-$V$)=0.2$. 

By constraining the colors to be within the range of the reddest MW and
bluest M31 globular
cluster colors (cf. the third selection criterion) I exclude distant
compact galaxies (Barrientos~\cite{barrientos}).
The total number of
{\it bona fide} globular cluster candidates is 71, with 
23 in Field~1 and 48 in
Field~2. Only those are taken into account in the further analysis.

\begin{figure*}
\centering
   \includegraphics[angle=270,width=17.5cm]{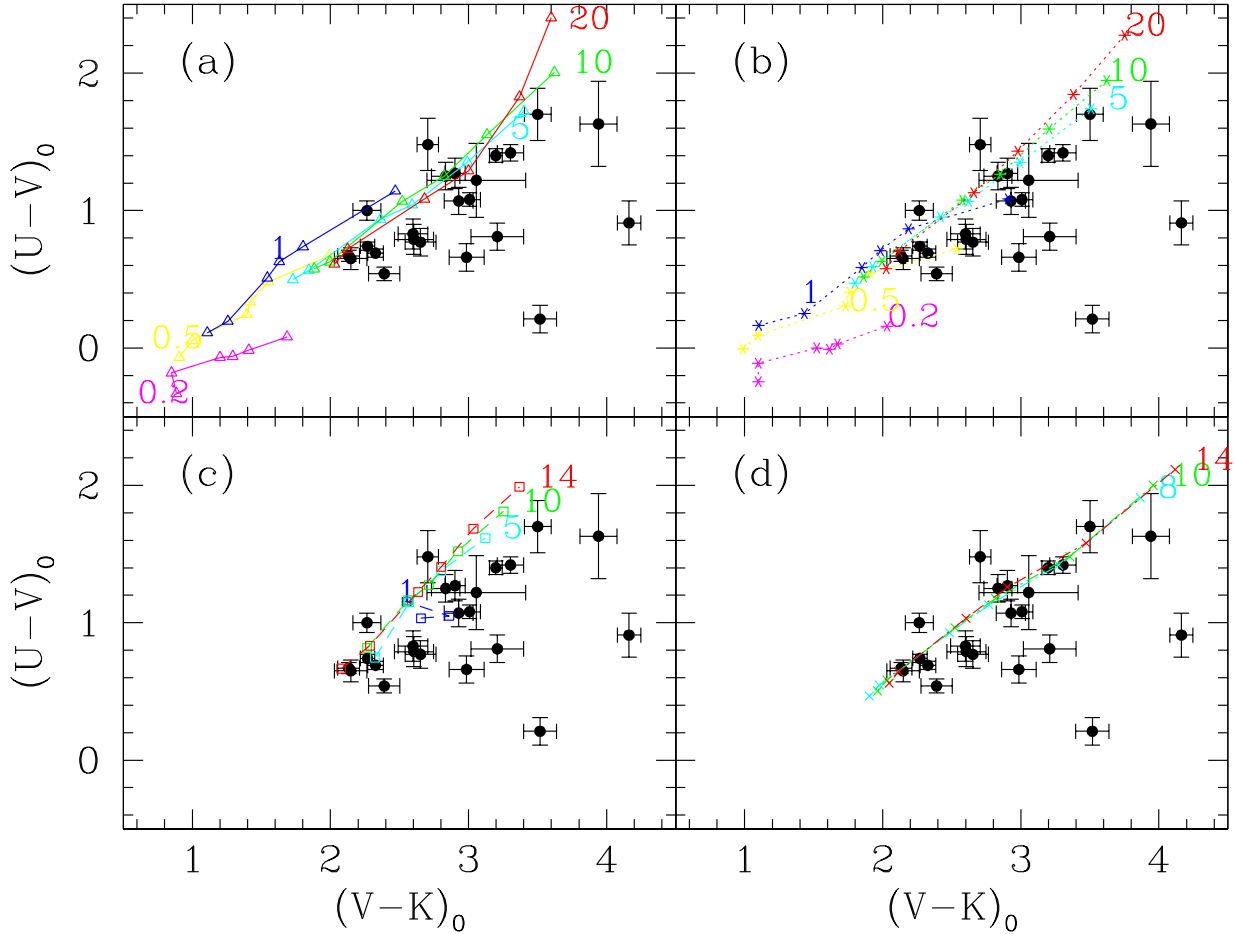}
   \caption[]{Color-color diagram for globular clusters
in NGC~5128 (filled circles). The lines are the isochrones of SSP models. 
The number next to each isochrone indicate the age in Gyr. Each isochrone is
plotted for a range of metallicities and for 
a Salpeter IMF. The four different models are from: 
(a) Kurth {\it et al.}\ (\cite{kurth}) for Z=0.0001, 0.0004, 0.008, 0.004,
0.02 and 0.05; (b) Bruzual \& Charlot (\cite{bc99}) for the same range of
age and Z
as the previous model; (c) Maraston (\cite{maraston00}) for [Fe/H]$=-2.25$ 
(only 14 Gyr isochrone), $-1.35$, $-0.55$, $-0.33$, 0.00 and 0.35 dex; 
and (d) Worthey
(\cite{wo94}) for Z=0.0002, 0.0004, 0.008, 0.004, 0.02 and 0.05. 
}
   \label{uvk}
\end{figure*}

K$_s$ band magnitudes were measured
on SOFI images for 23 of the {\it bona fide} globular cluster candidates 
and 10 of the extended objects,
some of which might also be globular clusters. 
The U$-$V color is strongly dependent on both metallicity and
age
(Buzzoni~\cite{buzzoni}; Kurth {\it et al.}\ \cite{kurth}; Maraston
\cite{maraston98}). 
Adding the K-band
magnitude information it is possible to disentangle the young from the old
objects. The color-color plot is shown in Fig.~\ref{uvk}. Filled dots are
used to denote globular cluster candidates that satisfied all of the 
classification criteria.
The lines are isochrones of evolutionary synthesis models
of simple stellar populations (SSP) calculated for a Salpeter IMF and a
range of metallicities from Z=0.0001 to 0.05 (see caption for the exact
metallicities plotted for each model). 
The four different models displayed on panels (a), (b), (c) and (d) are 
from Kurth {\it et al.}\ (\cite{kurth}),
Bruzual \& Charlot (\cite{bc99}), Maraston (\cite{maraston00}) and Worthey
(\cite{wo94}), respectively.
The ages are indicated with numbers denoting Gyr next to the lines. 
Unfortunately the models do not allow one 
to discriminate between ages in the range of
5-20 Gyr. However, it is comforting to see that almost all the objects are
concentrated towards the isochrones in excess of 1 Gyr. This is 
additional evidence that they are indeed globular clusters. 
A spread of
metallicities of more than 1.5 dex is also evident. 

In Fig.~\ref{uvk} most of the clusters lie slightly 
to the right and below the
model lines. There are two possible explanations. First, a
difference with the SSP models, particularly significant in
U-V and B-V colors, has already been pointed out by Barmby \& Huchra
(\cite{bh2000}). They suggested U fluxes from the
stellar libraries, systematic problems with modeling HB color and
observational errors as possible causes of this offset. Second, in the data
presented here additional offset may arise from the difference between the
Bessel and Johnson U-band transmission curves (Bessel \cite{bessel}).
 
\section{Luminosity Function}
\label{LF}

\begin{figure}
    \resizebox{\hsize}{!}{\includegraphics{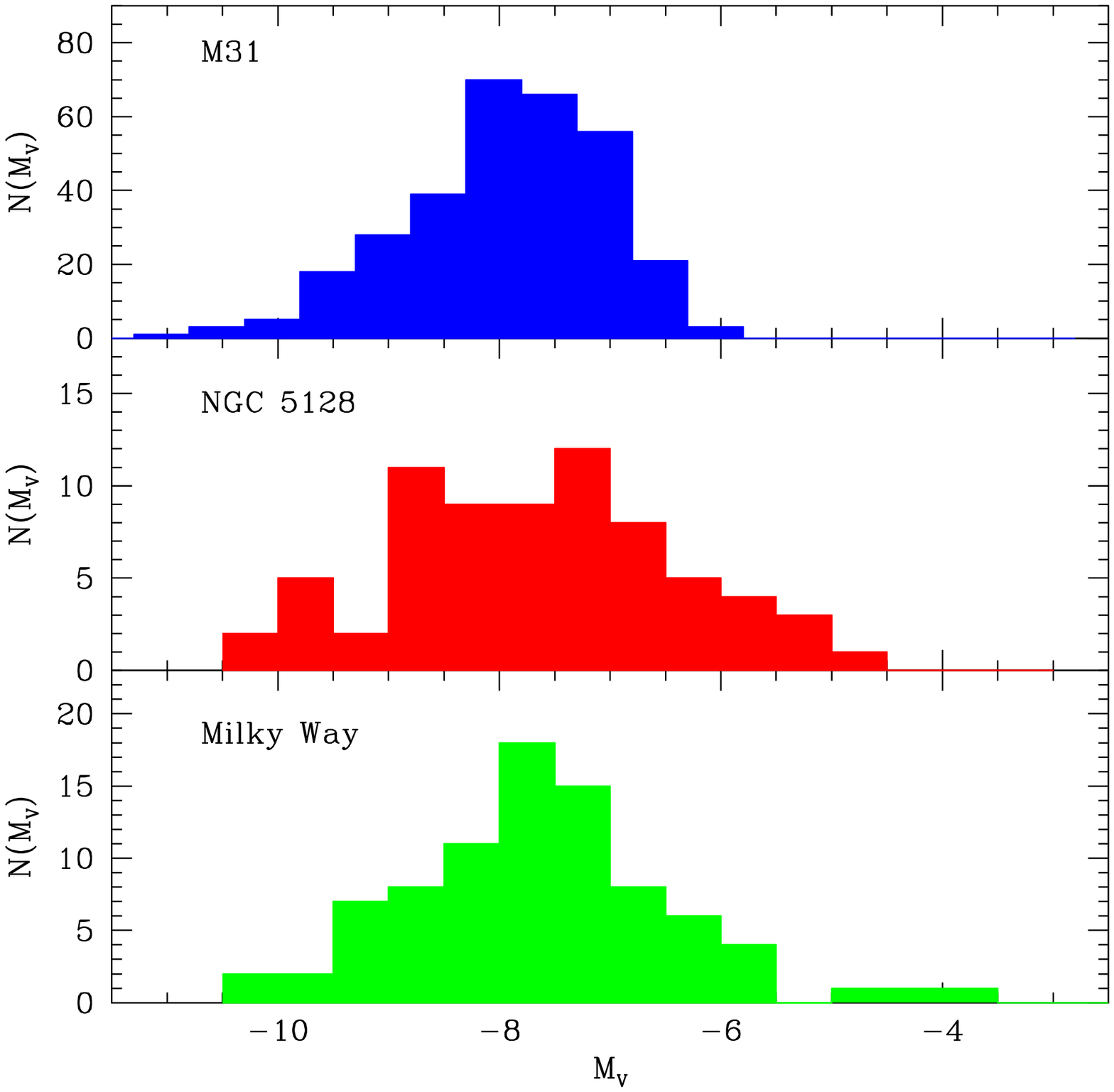}}
    \caption[]{From top to bottom: V-band GCLF for M31, NGC~5128 and MW. 
}
    \label{LFmag}
\end{figure}

\begin{figure}
    \resizebox{\hsize}{!}{\includegraphics[angle=270]{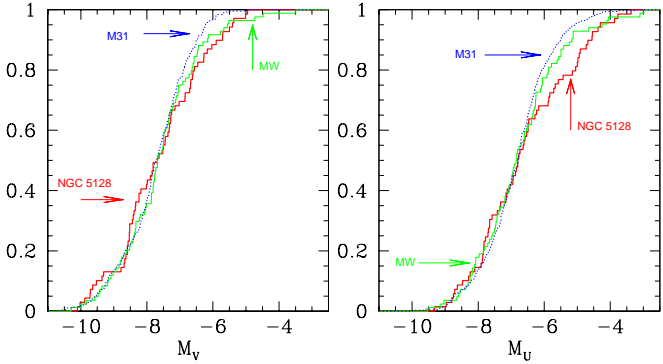}}
    \caption[]{Kolmogorov-Smirnov statistic: The cumulative distribution
function for NGC~5128 (thick line) is constructed from the V-band 
GCLF (left panel) and U-band GCLF (right panel) and compared
with the ones of the MW (thin line) and M31 (dotted line). 
}
    \label{kolmogorov}
\end{figure}

The V-band GCLF for NGC~5128 is compared to the ones of the MW and M31 in
Fig.~\ref{LFmag}. The actual shape of the luminosity function depends
on the size of the bins and on the bin-centers.
Statistically it is possible to compare them independently of
binning using the Kolmogorov-Smirnov (KS) test. 
In Fig.~\ref{kolmogorov} the cumulative distribution
functions for the three GCLFs are shown. The thick line is used for
NGC~5128, thin line for the MW and dotted line for M31.
Absolute magnitudes for NGC~5128 globular clusters were calculated using the
distance modulus value of (m$-$M$)_0=27.8$ (Soria {\it et al.}\ ~\cite{soria})
and reddening value of E(B$-$V$)=0.1$.

The KS statistic measures the 
maximum value of the absolute difference 
between two cumulative distribution functions from which a probability
that the two distributions are drawn from a common distribution 
can be calculated.
The probability that the GCLF of NGC~5128 matches the
one of the MW is 89\%, 
comparable to the probability that M31 GCLF matches the MW
one (85\%). The difference is slightly bigger in the U-band distributions
(Fig.~\ref{kolmogorov} right panel), but it is also probable that the
three luminosity functions are essentially similar for the three
galaxies. I conclude that the NGC~5128 GCLF is as similar (or as different)
to the MW GCLF as the M31 one.

Usually the GCLF is fitted with a gaussian or $t_5$ distribution 
function
in order to determine its peak magnitude. It should be noted, however, that 
the reasonably complete luminosity functions 
are known only for the two spiral galaxies, the MW and M31. 
Both of these GCLFs appear to depart from the gaussian
distribution (Ashman {\it et al.}\ \cite{acz}), while there is no complete
GCLF for any elliptical galaxy. The dependence of the turn--over magnitude
on the detection of the faint end of GCLF is discussed in Ashman {\it et
al.}\ (\cite{acz}). 
The determination of the turn--over magnitude
for NGC~5128 is left for a subsequent paper where the complete analysis of
WFI data is going to be presented. 

The detection of the faint end of the GCLF in an elliptical galaxy is also
important from the dynamical point of view. 
The faint globular clusters are
typically also the less massive ones and their 
presence in the deep potential
of a giant elliptical puts strong constraints on the potential itself as
well as on the effects of tidal forces in the halo. Unfortunately the
faintest globular clusters, 
corresponding to the MW Pal~1, AM~4, Terzan~1, Pal~13
are probably not detectable on FORS1 images (see Sect.~\ref{completenesstest}).

The processes that destroy globular clusters can be divided into internal,
like two-body relaxation and mass loss during stellar evolution, and
external. The external processes depend greatly on environment and therefore
on the position of the clusters in the galaxy. While tidal shocks
preferentially destroy less dense clusters, dynamical friction brings more
massive clusters very close to galactic centre. Both of these external
processes are much stronger in the inner part of the galaxy. 
According to theoretical simulations (Vesperini~\cite{vesperini}, Ostriker \&
Gnedin \cite{ostriker_gnedin}) in a galaxy
like NGC~5128, 
with the total mass of $\sim4\times10^{11}$ M$_\odot$ (Israel~\cite{israel})
and R$_e=5.24$ kpc (Dufour {\it et al.}\ \cite{dufour}), a strong
disruption and mass loss are efficient only within a galactocentric distance
of $1-2$ R$_e$, while at larger radii most of the mass loss is due to effects
of stellar evolution. Field~1 is located at $2-3.5$ R$_e$ and
Field~2 at $1.2-2.3$ R$_e$. While the dynamical effects may still be
important for Field~2 clusters, they are probably completely negligible for 
clusters in the outer field. It is not possible to quantify these effects by
simple comparison of the two subsamples, because of the small number
statistics.

\section{Color and Metallicity Distributions}
\label{color}

\begin{figure}
    \resizebox{\hsize}{!}{\includegraphics[angle=270]{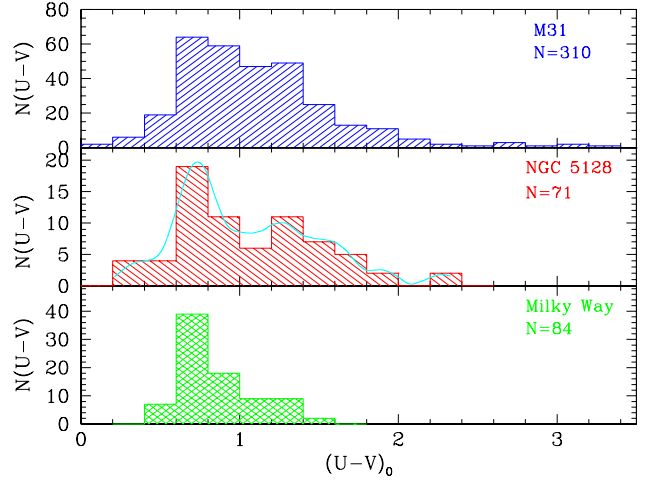}}
    \caption[]{The color distribution for M31, NGC~5128 and the MW. The line
overplotted on the NGC~5128 histogram is the best nonparametric 
kernel estimate (Silverman~\cite{silverman}) of the color distribution.}  
    \label{LFcolor}  
\end{figure}

The (U$-$V)$_0$ color histogram for 71 globular clusters in NGC~5128
indicates a bimodal distribution (Fig.~\ref{LFcolor}). 
Note that the (U$-$V)$_0$ color distribution of the MW globular
clusters does not show bimodality while the one of M31 does. The (U$-$V)$_0$
color distribution of the MW globular clusters is in fact biased due to lack
of U-magnitude and higher reddening values (only clusters with
E(B$-$V$)<0.5$ are plotted) of bulge (more metal-rich and redder) cluster. 
The blue peaks of the MW, M31 and NGC~5128 globular cluster 
color distributions
coincide. The blue cutoff in the three galaxies
is quite sharp and is driven principally by the colors of blue
horizontal branch stars. The red side of the color distribution is much
broader. In M31 it can be partially driven by differential reddening inside
the host spiral galaxy, because the colors for M31 clusters presented 
here were corrected only
for the mean foreground reddening of E(B$-$V)=0.08. Examination of
the
stellar (U$-$V) {\it vs.}\ (V$-$K) diagram in the two fields in NGC~5128
indicates that the amount of reddening does not vary much in the halo of
this giant elliptical (Rejkuba {\it et al.}\ \cite{rejkuba2}). 
Harris
{\it et al.}~(\cite{hetal92}) and Holland {\it et al.}~(\cite{holland})
also found more red (metal-rich) clusters in their NGC~5128 data.

The color distribution of the MW globular clusters in Fig.~\ref{LFcolor},
being dominated by halo clusters, is representative of the 
GCS of our Galaxy as if observed from outside. The NGC~5128 clusters
in Harris {\it et al.}~(\cite{hetal92}) and in this work belong as
well to the halo population. The differences in the color (metallicity)
distribution of clusters and stars in the halo of a spiral, like MW 
and a gE like NGC 5128, are reflected in figure 10 with a much larger
metallicity spread in gE galaxy.

The line overplotted on
the color distribution of NGC~5128 
is the best nonparametric kernel estimate (Silverman \cite{silverman}) of the
color distribution used to measure accurately the blue and the red peaks of
the distribution. The blue and red peaks are measured 
at (U$-$V)$_0=0.73\pm0.01$ and $1.28\pm0.01$, respectively.

\begin{figure}
    \resizebox{\hsize}{!}{\includegraphics[angle=270]{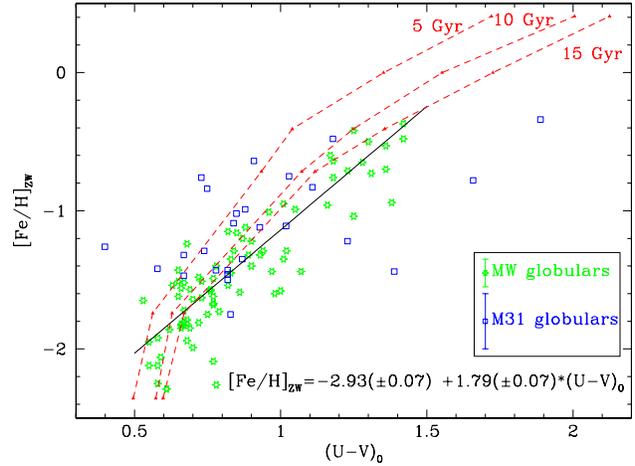}}
    \caption[]{The (U$-$V)$_0$ {\it vs.}\ [Fe/H] color-metallicity linear relation (full
line) derived from the MW globular clusters with E(B$-$V)$<0.5$ (stars).
The M31 clusters with spectroscopic metallicities and accurate (U-V) colors
plotted for comparison (squares) were not used for the fit. Typical errors
in measured metallicities are plotted on in the box.
Short dashed lines are the SSP models from Kurth {\it et al.}\
(\cite{kurth})
for ages of 5, 10 and 15 Gyrs, from blue to left, respectively. 
Each model spans metallicities from Z=0.0001 to 0.05.}
    \label{Fefit}     
\end{figure}

From the color-color diagram (Fig.~\ref{uvk}) it is obvious that the
globular clusters in NGC~5128 are older than 1 Gyr. Unfortunately, the
models do not allow one 
to better constrain the ages (Fig.~\ref{uvk}; see
also Barmby \& Huchra \cite{bh2000}).
Assuming that the globular clusters in NGC~5128 have similar ages as
the MW ones, one can use the Galactic globular clusters to calibrate the
metallicity scale. It is known that for the very low and very high
metallicities the
color-metallicity relation cannot be expressed by a simple linear function,
but extrapolation from lower metallicities towards the higher ones by using
the higher polynomial fit might lead to unphysical results. Thus I prefered
to make a linear fit with gaussian errors in $\sigma(Fe/H)=0.1$ in order
to calibrate the 
metallicity as a function of intrinsic (U$-$V)$_0$ color. Only
MW globular clusters from W.\ Harris (\cite{harris96})
web list with E(B$-$V)$<0.5$
(equivalent to E(U$-$V)$<0.82$; Rieke \& Lebofsky \cite{rl85})
were used
in the fit. The following color-metallicity relation is based on the Zinn \&
West (\cite{ZW}) scale and is valid only for the range of metallicities
observed in the MW globular clusters (Fig.\ ~\ref{Fefit}):
\begin{equation}
[Fe/H] = -2.93 (\pm 0.07) + 1.79 (\pm 0.07) \times (U-V)_0
\label{Fefit.equ}
\end{equation}

In the Fig.~\ref{Fefit} linear (U$-$V)$_0$ {\it vs.}\ [Fe/H] color-metallicity 
relation (Eq.~\ref{Fefit.equ}; full line) 
is plotted along with the evolutionary synthesis models 
of simple stellar populations (short dashed lines) 
from Kurth {\it et al.}\ (\cite{kurth}). Using the empirical linear
calibration and the SSP models I calculate the metallicity of the two peaks
in the color-distribution. The blue and the red peak 
correspond to [Fe/H]$=-1.7$ dex and [Fe/H]$=-0.6$ dex using the
linear relation (Eq.~\ref{Fefit.equ}) 
and to [Fe/H]$=-1.7$ dex and [Fe/H]$=-0.5$ dex using the models.

Although it is not obvious from the (U$-$V)$_0$ color distribution, but
rather from the spectroscopically measured [Fe/H] distribution, the
metallicity distribution of MW globular clusters is bimodal with the
metal-poor and metal-rich peaks at -1.6 and -0.6 dex respectively
(Harris~\cite{harris2000}). These values are practically the same as the
ones derived here for globular clusters 
in NGC~5128 from the (U$-$V)$_0$ colors.
On the other hand, on the basis of Washington photometry of 62 globular
clusters in NGC~5128, Harris {\it et al.}~(\cite{hetal92}) conclude that
{\it ``in both the range and distribution of abundance, the NGC~5128 GCS is
clearly different from the Galactic GCS and quite similar to that of the
large elliptical NGC~1399''}. They measured a mean metallicity of 
$\langle{\rm [Fe/H]}_{C-T_1}\rangle=-0.8\pm0.2$ dex, 
substantially more metal-rich than the
mean value for the MW GCS. They, however, mention that the values exceeding
[Fe/H]$\sim -0.25$ are extrapolated, a fact that is also true in the
metallicity calibration from (U$-$V)$_0$ color distribution. 
The extrapolation leads to large errors and a possible overestimation of
abundances of the metal-rich clusters. 
Because of this I do not report here the metallicities of individual
clusters. 

Note also that I restricted the (U$-$V)$_0$ color to be 
within the range observed in MW and bulk of the M31 globular
clusters. In this way the sample was selected to be as clean as 
possible from the very compact and red galaxies.
It is not surprising that the clusters with
supra-solar abundances are not present, 
reducing the mean abundance of the sample. 
However, relaxing the red color cut
and adding the 6 ``good'' globular cluster candidates redder than
(U$-$V)$_0=2.5$ (filled symbols in Fig.~\ref{CMDgc.CenA}) 
the mean of the color
distribution shifts from $\langle{\rm (U-V)_0}\rangle=1.05\pm0.45$ to
$\langle{\rm (U-V)_0}\rangle=1.22\pm0.75$, corresponding to mean metallicities
of $\langle{\rm [Fe/H]}\rangle=-1.1\pm0.1$ dex and $\langle{\rm
[Fe/H]}\rangle=-0.7\pm0.2$ dex, respectively. The latter value is very
similar to the mean metallicity derived from Washington photometry by Harris
{\it et al.}~(\cite{hetal92}).
In order to assess the real nature of the objects redder than (U$-$V)$_0=2.5$
spectroscopic observations are necessary.

\section{Conclusions}
\label{conclusion}

The main results and their implications 
presented in this work are the following:

\begin{itemize}

\item[1.] On the basis of high resolution ground-based images, taken with VLT
and FORS1, I identified 71 {\it bona fide} globular clusters in the halo of the
nearest giant elliptical galaxy NGC~5128. Only 5 of these clusters were
previously known. 

\item[2.] In a UV color-magnitude diagram for the complete sample of 
71 clusters and a UVK
color-color diagram for a subsample of 23 clusters, the objects span a 
similar magnitude and color range as the globular clusters in the MW and
M31. Note, however, that there is no cluster corresponding to $\omega$ Cen,
the brightest MW cluster, in these two fields. 
The position of 23 objects in the (U$-$V)$_0$ {\it vs.}\ (V$-$K)$_0$
color-color diagram is consistent with their classification as old  
globular clusters.

\item[3.] The GCLFs spanning $-10.1<M_V<-4.9$
and $-9.3<M_U<-3.3$ have been constructed. These are the deepest GCLFs of an
elliptical galaxy made so far. 
Kolmogorov-Smirnov statistics show that the
difference between the GCLFs of NGC~5128 and MW is not larger than 
the difference between the GCLFs of M31 and MW. Similarity of the GCLFs of
an elliptical with respect to the spiral galaxy had never been tested at the
faint end before.

\item[4.] The presence of faint globular clusters in the halo of
NGC~5128 puts constraints on the effectiveness of the tidal forces in the
deep elliptical galaxy potential. The dynamical effects may be
important for clusters that are found within $\sim2$ R$_e$ from the galactic
center. Unfortunately, the selection of globular
clusters from VLT images is not sensitive to the most compact and the
faintest clusters, similar to ones like Pal~1, Pal~13, AM~4 and
Terzan~1 in our Galaxy, 
while the less dense ones start to be confused with
background galaxies at faint magnitudes.

\item[5.] The (U$-$V)$_0$ color histogram of 71 clusters indicates 
a bimodal distribution, supporting 
the Zepf \& Ashman ~(\cite{zepfashman93})
suggestion. Assuming that the clusters in NGC~5128 and in the MW span a
similar age interval, and adopting the linear fit between the (U$-$V)$_0$
color and metallicity, I derived the [Fe/H] of the red and blue peaks of the
bimodal distribution to be $-1.7$ dex and $-0.6$ dex, respectively. Using
the SSP models from Kurth {\it et al.}\ ~(\cite{kurth}), instead of the
linear fit, the results do not change significantly, giving values of
$-1.7$ dex and $-0.5$ dex for the metal-poor and 
metal-rich peaks, respectively. This
is different from the Harris {\it et al.}~(\cite{hetal92}) and Zepf \&
Ashman ~(\cite{zepfashman93})
result, but is partially due to the adopted color cut of 
(U$-$V)$_0<2.5$ in selecting globular clusters 
and to small number statistics, since both
samples have $<10$\% of the total cluster population. 
Relaxing the red color cut the mean color of the
distribution corresponds to mean metallicity of
[Fe/H$]=-0.7$ dex, very similar to Harris {\it et
al.}~(\cite{hetal92}) result.
The true nature of the redder objects has to be assessed
through spectroscopy.

\end{itemize}

\begin{acknowledgements}
I am grateful to Dante Minniti for his guidance and help with this project.
I thank Andres Meza for providing the code that calculates the kernel
estimator. I would like to acknowledge the referee for valuable 
comments which led to improvements in the paper and useful discussions with 
Markus Kissler-Patig, Thomas Puzia and Steve Zepf. Thanks also go to Tim
Bedding, Dave Silva and Elena Pancino.
This research was partially supported
by FONDECYT grant No.\ 1990440.

\end{acknowledgements}

\end{document}